\documentclass{SciPost}

\hypersetup{
    colorlinks,
    linkcolor={red!50!black},
    citecolor={blue!50!black},
    urlcolor={blue!80!black}
}

\usepackage[bitstream-charter]{mathdesign}
\DeclareSymbolFont{usualmathcal}{OMS}{cmsy}{m}{n}
\DeclareSymbolFontAlphabet{\mathcal}{usualmathcal}

\fancypagestyle{SPstyle}{
\fancyhf{}
\lhead{\colorbox{scipostblue}{\bf \color{white} ~SciPost Physics }}
\rhead{{\bf \color{scipostdeepblue} ~Submission }}

\fancyfoot[C]{\textbf{\thepage}}
}

\begin{document}

\pagestyle{SPstyle}

\begin{center}{\Large \textbf{\color{scipostdeepblue}{
%%%%%%%%%% TODO: Write your article's title here
Lattice polarons with extended interactions\\
}}}\end{center}

\begin{center}\textbf{
% Use (full) first name (+ middle name initials) + surname format.
% Separate subsequent authors by a comma, omit comma and use "and" for the last author.
% Mark the corresponding author(s) with a superscript symbol in this order
% \star, \dagger, \ddagger, \circ, \S, \P, \parallel, ...
Enrique I. Ramírez-Juárez\textsuperscript{1$\star$},
Genaro Lopez-Olivera\textsuperscript{1$\star$},
L.~A.~Pe\~na Ardila\textsuperscript{2,3} and
Arturo Camacho-Guardian\textsuperscript{1$\dagger$}
}\end{center}

\begin{center}
{\bf 1} Instituto de Fisica, Universidad Nacional Autonoma de Mexico, Ciudad de Mexico CP 04510, Mexico
\\
{\bf 2} Dipartimento di Fisica, Universit\`a di Trieste, 
Strada Costiera 11, I-34151 Trieste, Italy
\\
{\bf 3} Istituto Nazionale di Fisica Nucleare (INFN), 
Trieste Section, Via Valerio 2, I-34127 Trieste, Italy
%%%%%%%%%% END TODO: AFFILIATIONS
% Provide email address of corresponding author(s)
\\[\baselineskip]

$\dagger$ \href{mailto:acamacho@fisica.unam.mx}{\small acamacho@fisica.unam.mx}
\end{center}

\section*{\color{scipostdeepblue}{Abstract}}
\textbf{\boldmath{%
Lattice impurities have recently emerged as a platform in which polarons unveil new quantum many-body states absent in free space and can serve to probe strongly correlated matter. In this work, we investigate two-dimensional lattice polarons with strong on-site repulsion and tunable nearest-neighbor interactions using a  variational approach  including up to one  excitation of the medium. We show that extended interactions qualitatively modify the quasiparticle structure beyond the conventional attractive and repulsive polaron picture. A direct analysis of the eigenvalue spectrum reveals the presence of dark impurity states, orthogonal to the bare impurity and therefore spectroscopically dark. These states exhibit nontrivial internal structure, including dipolar symmetries in real space. Our results demonstrate that long-range interactions generate  multiple quasiparticle excitations with distinct symmetry properties, highlighting the crucial role of interaction range and lattice geometry. This work opens new avenues for probing hidden quasiparticle states in lattice systems through spectroscopic and wave-function-resolved measurements.
}}

\vspace{\baselineskip}

%%%%%%%%%% BLOCK: Copyright information
% This block will be filled during the proof stage, and finilized just before publication.
% It exists here only as a placeholder, and should not be modified by authors.
\noindent\textcolor{white!90!black}{%
\fbox{\parbox{0.975\linewidth}{%
\textcolor{white!40!black}{\begin{tabular}{lr}%
  \begin{minipage}{0.6\textwidth}%
    {\small Copyright attribution to authors. \newline
    This work is a submission to SciPost Physics. \newline
    License information to appear upon publication. \newline
    Publication information to appear upon publication.}
  \end{minipage} & \begin{minipage}{0.4\textwidth}
    {\small Received Date \newline Accepted Date \newline Published Date}%
  \end{minipage}
\end{tabular}}
}}
}
%%%%%%%%%% BLOCK: Copyright information

% For convenience during refereeing we turn on line numbers:


\vspace{10pt}
\noindent\rule{\textwidth}{1pt}
\tableofcontents
\noindent\rule{\textwidth}{1pt}
\vspace{10pt}

% Write your article contents here, starting from first \section.
% An example structure is given below.

%==========================================================
\section{Introduction}
The problem of a mobile impurity coupled to a quantum bath remains a powerful framework for understanding quasiparticle formation in many-body states. The concept of the polaron has extended from Landau's and Pekar's paradigm~\cite{landau1948effective} to recent experiments in several platforms, where the ability to tune the scattering into the strongly interacting regime via Feshbach resonances in quantum gases~\cite{Chin2010} and bi-excitons in semiconductors~\cite{carusotto2010feshbach,Kuhlenkamp2022} has unveiled new classes of Bose polaron states in ultracold gases~\cite{Jorgensen2016,Hu2016,yan2020bose,Ardila2019,skou2021non,Etrych2025}, polaron-polaritons~\cite{takemura2014polaritonic,navadeh2019polaritonic,takemura2017spin,Tan2020,Li2021,Bastarrachea2019,Tan2023,Choo2024}, among others~\cite{massignan2025polarons,grusdt2025impurities}.

Most of the progress has been achieved in the context of polarons in continuum systems. However, the situation in lattice systems leads to additional ingredients, such as a discrete band structure and reduced symmetry, which can give rise to new channels for quasiparticle dressing.

Lattice polarons have  recently gained attention as they have been proposed as probing schemes for topological phases~\cite{grusdt2016interferometric,Grusd2019,Camacho2019,Pimenov2021,vashisht2025chiral,wagner2026sensing,Julia2020,Baldelli2021,Alhyde2022} and the superfluid-Mott transition~\cite{Alhyde2022,alhyder2025lattice}. Furthermore, lattice Bose polarons have revealed intriguing states such as stable repulsive polarons and bipolarons~\cite{Isaule2025,ding2023polarons,dominguez2023bose,Felipe2024,isaule2026mobile,Yordanov_2023,rojo2024few,Hartweg2025,zhang2026fate,Lozada2025}, Rydberg lattice polarons~\cite{Castro2026}, and the polaron in a condensate of hard-core bosons~\cite{santiago2024lattice,santiago2026moire}. In most of these studies, the coupling between the impurity and the medium is assumed local. Long-range interactions arise naturally in platforms where dipolar interactions are intrinsic to the impurities, for instance, in two-dimensional multi-layer heterostructures~\cite{li2020dipolar,sun2024dipolar,erkensten2023electrically,louca2023interspecies,herrera2025moire}. These long-range interactions can lead to new quantum many-body states absent in purely local models~\cite{Julku2022,deng2025frozen}. Despite their relevance, the impact of extended interactions on the structure of lattice polarons is still poorly understood. 

In this work, we investigate a lattice polaron with strong on-site repulsion and a tunable nearest-neighbor interaction. By employing a  variational ansatz including up to one excitation, also known as Chevy variational ansatz, we compute the eigenstates and the spectral function of the system. We demonstrate that, in contrast to the conventional picture~\cite{ding2023polarons,dominguez2023bose,Lozada2025}, the system supports a robust set of stable repulsive polaron branches. By directly analyzing the eigenvalue spectrum, we demonstrate the presence of dark impurity states, orthogonal to the bare impurity state and therefore not accessible in the spectral function.

\begin{figure}[t]
    \centering
    \includegraphics[width=0.45\columnwidth]{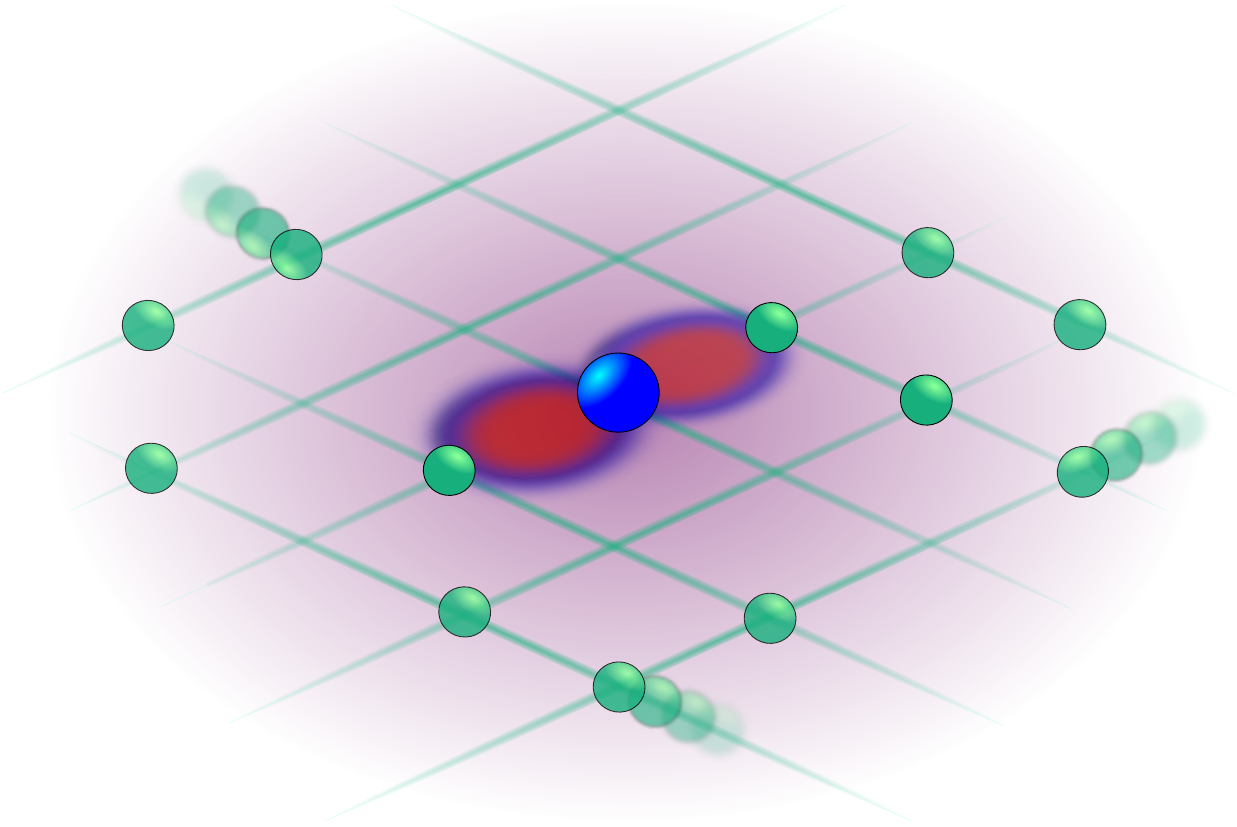}
    \caption{
    Schematic illustration of a lattice polaron in a two-dimensional optical lattice. The central blue sphere represents the impurity, while the surrounding green spheres denote the bosonic medium localized on lattice sites. Non-local interactions can lead to dark impurity states. }
    \label{fig:cartoon}
\end{figure}

Our study establishes that extended interactions generate several quasiparticle excitations with nontrivial symmetry and internal structure, highlighting the importance of lattice geometry and interaction range in polaron physics. Furthermore, the intricate spatial structure induced by nonlocal interactions may provide a route toward exotic few- and many-body polaron phases, including spatially nontrivial bipolarons and density-wave states.

%==========================================================
\section{System}
We consider an impurity confined in a square two-dimensional optical lattice coupled to an ideal Bose gas, as illustrated in Fig.~\ref{fig:cartoon}. The coupling between the impurity and the bosons accounts for a strong on-site repulsion and a tunable nearest-neighbor interaction, such that the Hamiltonian reads as,
\begin{gather}
\hat H=-t_b\sum_{\langle i,j\rangle}\hat b_{\mathbf r_i}^\dagger\hat b_{\mathbf r_j}-\sum_{i}\mu_B\hat b^\dagger_{\mathbf r_i}\hat b_{\mathbf r_i}-t_c\sum_{\langle i,j\rangle}\hat c_{\mathbf r_i}^\dagger\hat c_{\mathbf r_j}+ \\ \nonumber
+U_0\sum_{i}\hat b_{\mathbf r_i}^\dagger\hat c_{\mathbf r_i}^\dagger \hat c_{\mathbf r_i} \hat b_{\mathbf r_i}+U_I\sum_{\langle i,j\rangle}\hat b_{\mathbf r_i}^\dagger\hat c_{\mathbf r_j}^\dagger \hat c_{\mathbf r_j} \hat b_{\mathbf r_i},
\end{gather}
where $\hat b_{\mathbf r_i}^\dagger/\hat c_{\mathbf r_i}^\dagger$ creates a boson/impurity at site $i$. The first line accounts for the hopping of the bosons  with tunneling amplitudes $t_b/t_c$ respectively and the chemical potential (first two terms), whereas the last term describes the tunneling of the impurity. We consider nearest-neighbor tunneling such that the dispersion of the bosons/impurity is simply given by 
$$\epsilon^{(b,c)}_{\mathbf{p}}=-2t_{b,c}[\cos(p_x a)+\cos(p_y a)],$$ with $\boldsymbol{p} = (p_x, p_y)$ a vector in the first Brillouin zone and $a$ the lattice constant. We consider a zero-temperature BEC and fix the chemical potential to the band minimum $\mu_B=\epsilon^{(b)}_{\mathbf{0}}$. In our model, we consider an on-site interaction $U_0$ and a first-neighbor interaction $U_I$. To understand the effects of the latter, in the following we consider a strong on-site repulsion, whereas $U_I$ is tuned.

\subsection{Two-Body States} Before we delve into the complexity of the many-body problem, it is illustrative to solve the two-body scattering problem between the impurity and a boson from the medium. We study the two-body problem for a fixed center-of-mass momentum, taking $\mathbf{K}=0.$ In this sector, the basis consists of states where the impurity and boson carry different momenta $|\text{TB}\rangle=\sum_{\mathbf{p}}\mathcal{B}_{\mathbf{p}}\hat{b}_{-\mathbf{p}}^\dagger\hat{c}_{\mathbf{p}}^\dagger|0\rangle_b\otimes|0\rangle_c.$ The two-body problem 
\begin{gather}
    \left(\epsilon^{(b)}_{\mathbf{k}}+\epsilon^{(c)}_{\mathbf{k}}\right)\mathcal B_{\mathbf{k}} 
+ \frac{1}{N}\sum_{\mathbf{k}'} V(\mathbf{k}'-\mathbf{k})\mathcal B_{\mathbf{k}'} 
 = E_{\text{2B}} \mathcal B_{\mathbf{k}},
\end{gather} 
can be translated into a standard matrix eigenvalue problem. By discretizing the Brillouin zone and diagonalizing this matrix, we obtain the full set of eigenvalues $E_{\text{2B}}$ and eigenvectors $\mathcal{B}_n(\mathbf{k})$ that describe the two-body spectrum.
\begin{figure}[t]
    \centering
    \includegraphics[width=.75\columnwidth]{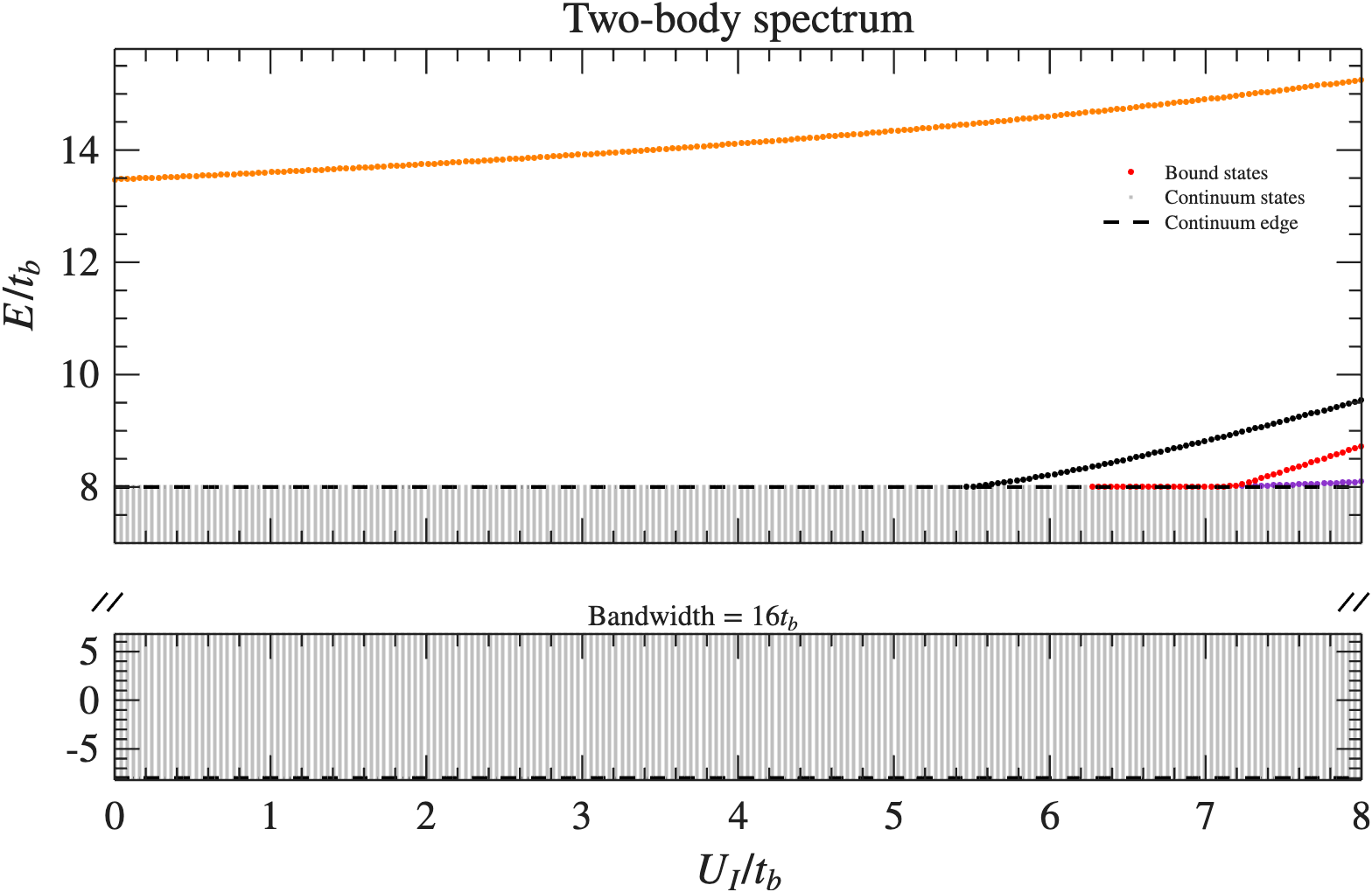}
    \caption{Scattering states of the two-body problem. The gray points correspond to scattered states. The orange dots correspond to a repulsive bound state from the on-site interaction whereas the black (two degenerate) and red dots correspond to secondary repulsive bound arising from the nonlocal interactions.
   }
    \label{fig:scattering}
\end{figure}

In Fig.~\ref{fig:scattering} we show the two-body scattering spectrum. For our numerical results, we take $t_b = t_c$ and a strong on-site interaction $U_0/t_b = 12$. We tune $U_I$. The gray region corresponds to the continuum of scattering states, with a bandwidth of $16t_b$. Above the continuum, the strong on-site interaction gives rise to a repulsive bound state marked by the orange points~\cite{winkler2006repulsively,valiente2008two,valiente2009scattering,valiente2010lattice,2y2n-1pkj}. This state lies far from the continuum and, since its energy exceeds that of any available scattering state, the pair cannot decay.  We take $N_{\text{sites}}=N_x\times N_y=80\times80=6400$ and have checked that our results are converged.

Close to $U_I/t_b\approx 5.49$, a pair of degenerate bound states emerge (black points). These states arise as a consequence of the non-local interaction, as they only appear for strong $U_I$. An additional bound state emerges close to $U_I/t_b \approx 7.11$, shown by the red dots and a weakly bound state barely on top of the continuum at $U_I/t_b \approx 8.00.$

The two-body scattering problem shows a repulsive bound state emerging from the on-site interaction, but it also reveals the presence of repulsive bound states that arise for strong non-local interactions.

\subsection{ Many-Body Ansatz} Having understood the two-body problem, we can now turn to the many-body problem. For this, we employ a variational ansatz restricted to one Bogoliubov excitation,
\begin{equation}
|{\psi}\rangle=\left(\varphi_0 \hat{c}^\dagger_{\mathbf{0}}+\sum_{\mathbf{p}}\psi_{\mathbf{p}}\hat{b}^\dagger_{-\mathbf{p}}\hat{c}^\dagger_{\mathbf{p}}\right)|{\mathrm{BEC}}\rangle\otimes|{0}\rangle_c,
\end{equation}
which describes the superposition of the bare impurity with an impurity dressed by a single excitation of the BEC. This ansatz is non-perturbative in the impurity-boson coupling and captures the relevant two-body physics to account for bound states. Our ansatz is the equivalent for bosons of the Chevy ansatz~\cite{Chevy2006}, originally proposed for the Fermi polaron.  Despite its apparent simplicity, this formalism has proven accurate in describing polaron experiments in the strongly interacting regime~\cite{Hu2016,Jorgensen2016,Ardila2019}. Here, $\hat{b}^\dagger_{\mathbf{p}}/\hat{c}^\dagger_{\mathbf{p}}$ are the field operators in momentum space.

By minimization of the variational energy $\langle \psi | \hat{H} - E | \psi \rangle$, we obtain a set of equations for $\varphi_0$ and $\phi_{\mathbf{k}}$

\begin{gather}
\left(\epsilon^{(c)}_{\mathbf{0}}+\frac{N_0}{N}V(\mathbf{0})\right)\varphi_0 
+ \frac{\sqrt{N_0}}{N}\sum_{\mathbf{q}} V(\mathbf{q}) \psi_{\mathbf{q}} = E \varphi_0, \\
\left(\epsilon^{(b)}_{\mathbf{k}}-\mu_B+\epsilon^{(c)}_{\mathbf{k}}\right)\psi_{\mathbf{k}} 
+ \frac{1}{N}\sum_{\mathbf{k}'} V(\mathbf{k}'-\mathbf{k})\psi_{\mathbf{k}'} 
+ \frac{\sqrt{N_0}}{N}V(-\mathbf{k})\varphi_0 = E \psi_{\mathbf{k}},
\end{gather}
here, $V(\mathbf k)=U_0+2U_I(\cos(k_x a)+\cos(k_y a))$ denotes the impurity-boson interaction accounting for both the on-site and long-range interactions.

The coupled equations can be recast as a matrix eigenvalue problem,
\begin{equation}
\hat{H}_{\mathrm{var}} \, \Psi = E \Psi,
\label{linearE}
\end{equation}
where $\Psi=(\varphi_0,\{\phi_{\mathbf{k}}\})^T$. Here, we obtain a set of $N_{\text{sites}} + 1$ eigenvectors and eigenvalues, which arise from discretizing the Brillouin zone and performing a numerical diagonalization using standard linear algebra routines. From the eigenvectors, we compute the spectral function as follows
\begin{equation}
A(\omega) = \pi\sum_n |\langle n | \psi_0 \rangle|^2 \, \delta(\omega - E_n),
\end{equation}
where $|\psi_0\rangle=|\text{BEC}\rangle\otimes|0\rangle_c$ denotes the noninteracting impurity state. For visibility purposes, we add a small imaginary part to the spectral function of $\eta=0.1t_b,$ that is, we take $\pi\delta(\omega-E_n)\rightarrow \eta/((\omega-E_n)^2+\eta^2)$

The variational approach employed here corresponds to a restriction of the full Hilbert space to states containing at most a single bosonic excitation on top of the condensate. The energies and wave functions capture the dominant two-body correlations between the impurity and the medium, in particular, it is a formalism well-suited for accounting for attractive and repulsive bound states~\cite{Rath2013,ding2023polarons}.

%The eigenstates that are orthogonal to $|\psi_0\rangle$ within this restricted Hilbert space naturally appear as dark states in the spectral function. While additional multi-excitation processes beyond the Chevy ansatz may further renormalize the spectrum, they are not expected to qualitatively alter the existence or symmetry properties of these states.

The Chevy ansatz is equivalent to the so-called T-matrix approach~\cite{Rath2013}, and both formalisms lead to the same spectral properties. However, as we will discuss below, in order to fully characterize the polaron, we also analyze the polaron wave function, which can be obtained directly by diagonalizing the effective Hamiltonian. In Appendix~\ref{Tmatrix} we discuss in detail the T-matrix and compare with the presented formalism.

%==========================================================
\section{Many-body  spectrum}

\begin{figure}[t]
    \centering
    \includegraphics[width=.75\columnwidth]{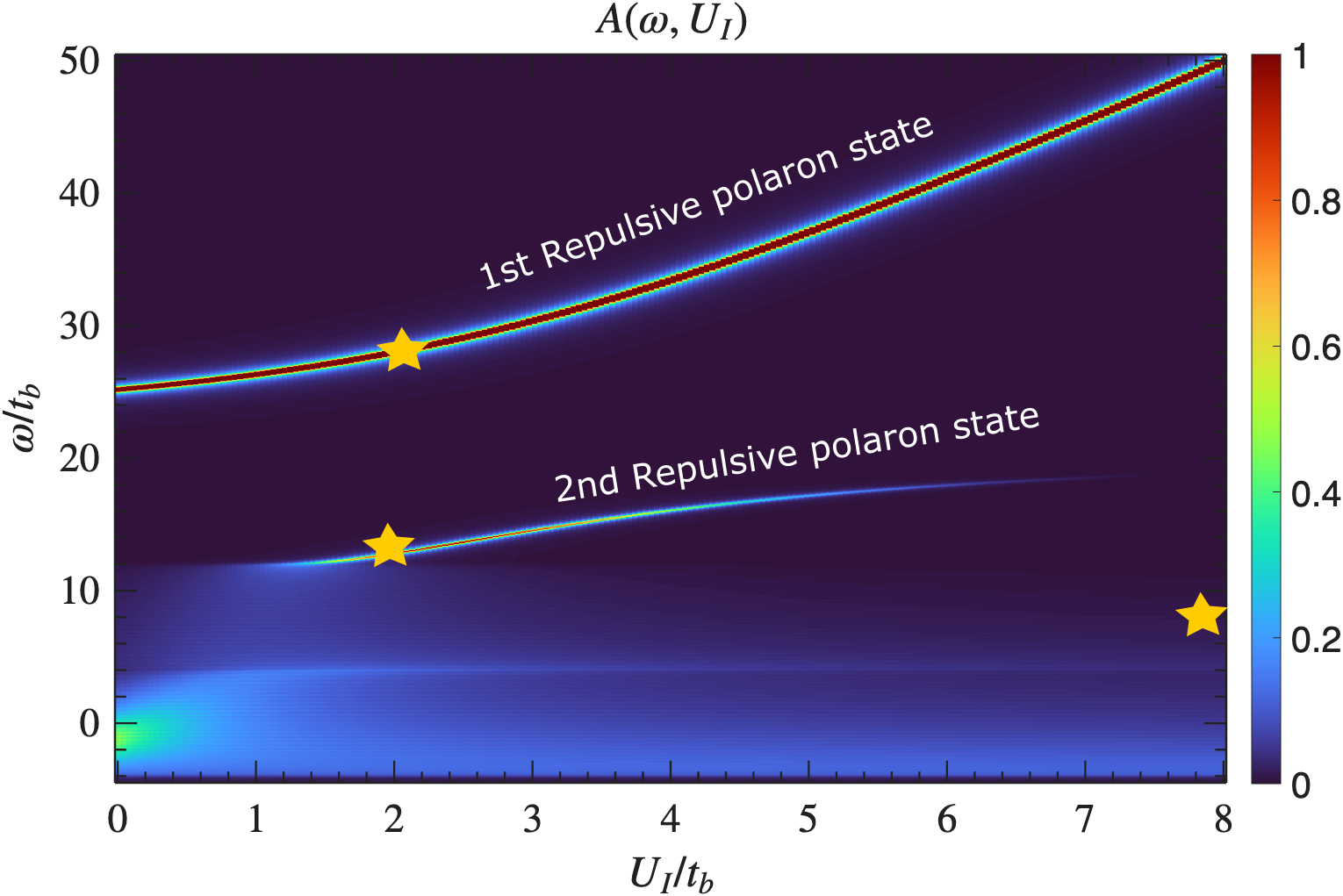}
    \caption{
    Spectral function $A(\omega)$ of a lattice polaron for fixed on-site interaction $U_0 = 12t_b$, shown as a function of frequency $\omega/t_b$ and nearest-neighbor interaction $U_I/t_b$. 
    }
    \label{fig:spectral}
\end{figure}
To start understanding the polaron states arising from the non-local interaction, we first study the spectral function. 

Fig.~\ref{fig:spectral} shows the spectral function as a function of $\omega$ as $U_I$ is varied. For small $U_I/t_b$, we find that the spectral function is distributed among two polaron states: a low-energy branch and a repulsive polaron branch. The former, which we refer to as the {\it lowest polaron state}, lies close to zero energy and sits just below the continuum of Bogoliubov excitations, which makes it diffuse in the spectral function. The {\it 1st repulsive polaron state} is a well-defined quasiparticle branch far from the continuum. This polaron branch arises from the coupling of the impurity to the highest repulsive bound state in Fig.~\ref{fig:scattering} (orange dots). For vanishing $U_I/t_b$, the interplay between these two branches has been previously studied in Ref.~\cite{ding2023polarons}. In this work, we focus instead on the new imprints that emerge from the non-local interaction.

%For small $U_I/t_b$ we find that the spectral function is mainly distributed among two polaron states: a) the attractive and b) repulsive branches. Here, the attractive polaron sits at an energy close to zero, and close to the continuum, which makes this branch diffuse. In contrast, at high energies, a well-defined repulsive polaron arises far from the continuum. As it has been discussed previously, this repulsive state is long-lived; it corresponds to a high-energy state where an impurity and a boson sit in the same lattice site. Since their energy is much larger than the energy of any scattering state available, it cannot break and is therefore stable.

%The interplay between the repulsive bound state and the low-lying polaron state in the lattice polaron has been discussed previously in Ref.~\cite{ding2023polarons}. Here instead we focus on the new imprints of the non-local interaction. The first feature is the appearance of an additional polaron branch.

As the non-local interaction increases, close to $U_I/t_b$, a third polaron branch arises, starting just above the continuum. This state, {\it 2nd repulsive polaron}, smoothly separates from the continuum as $U_I/t_b$ is increased. The state evolves from being ill-defined to gaining significant spectral weight and becoming well-defined around $U_I/t_b \sim 3$, to losing all visibility around $U_I/t_b > 6$. We should note that this state is absent for $U_I = 0$, that is, it is a consequence of the non-local interactions between the impurity and the bosons. 

\begin{figure}[t]
    \centering
    \includegraphics[width=0.75\columnwidth]{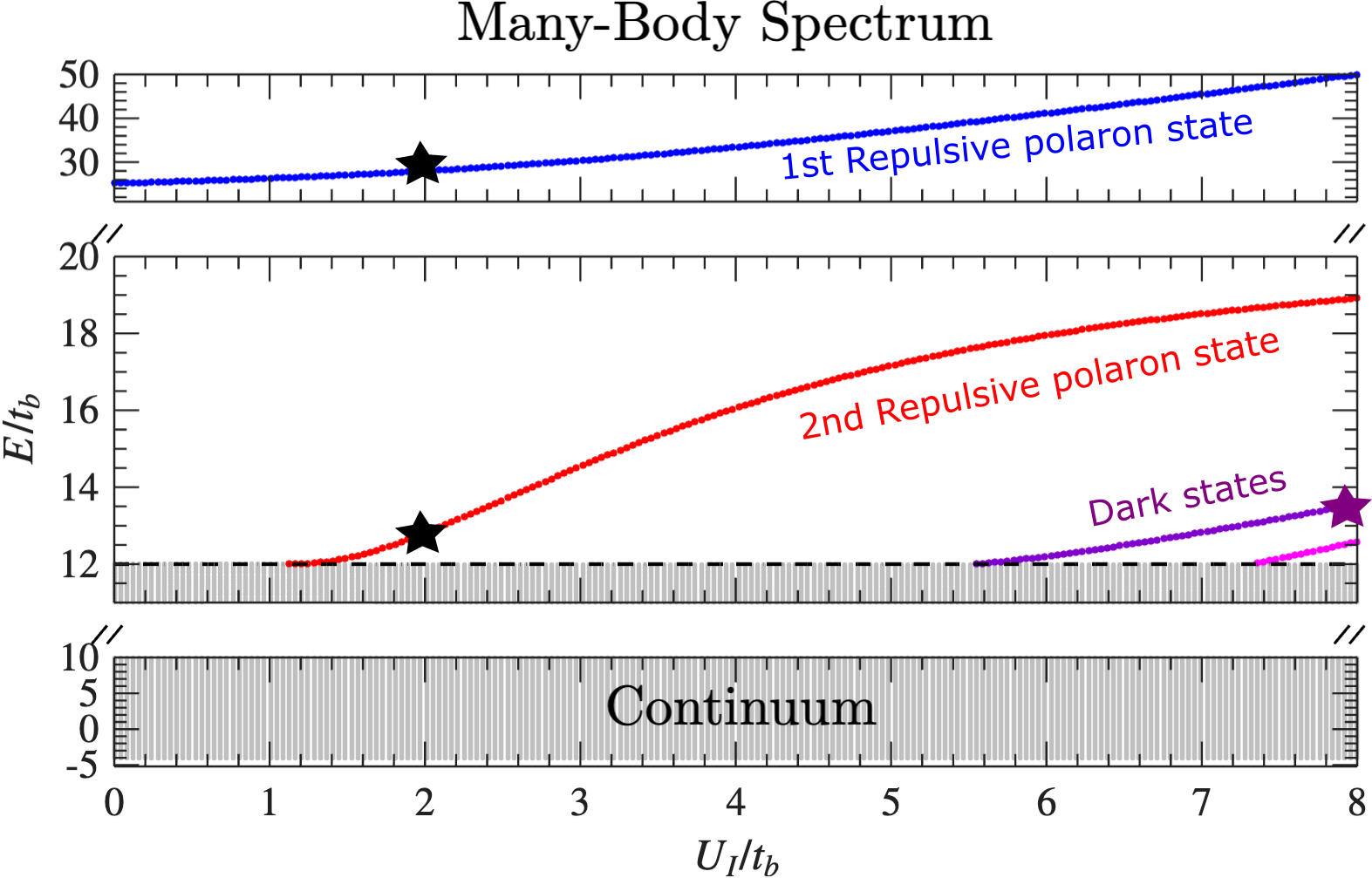}
    \caption{
    Eigenenergy spectrum obtained from the Chevy ansatz for a lattice polaron with fixed on-site interaction $U_0=12t_b$, as a function of nearest-neighbor interaction $U_I/t_b$. 
    }
    \label{fig:eigenstates}
\end{figure}

The spectral function already suggests that the structure of the many-body spectrum is richer than for solely on-site interactions. To further understand the character of the polaron states, let us analyze how the full energy spectrum of $\hat{H}_{\text{var}}$ looks as a function of $U_I/t_b$. This is shown in Fig.~\ref{fig:eigenstates}, where we can identify in blue and red the 1st and 2nd repulsive polaron states, which indeed, as the spectral function showed, are states separated from the continuum, well-defined, and long-lived. For $U_I/t_b \sim 7$, additional states separated from the continuum emerge. These spectroscopically dark state, arise as a consequence of the secondary repulsive bound states shown in the two-body scattering spectrum in Fig.~\ref{fig:scattering}.

Interestingly, the {\it 2nd repulsive polaron state} arises as a consequence of the interplay between the on-site and non-local interactions. However, the emergence of this branch is not accompanied by an additional bound state, as is the case for the {\it dark impurity} states appearing in the many-body spectrum.

On the other hand, the {\it dark-impurity states} are genuine many-body eigenstates of the Hamiltonian; however, they have a vanishing overlap with the bare impurity state. Although these states arise as eigenstates of the many-body problem, they have a clear two-body origin, the remnants of the bound states shown in Fig.~\ref{fig:scattering}. 

A very important implication of our results is that complex polaron states may remain invisible in conventional spectroscopic probes. In the present case, these states do not appear in the spectral function because they have vanishing overlap with the bare impurity state. Therefore, a detailed analysis of both the two-body and many-body spectra is required to obtain a complete understanding of the problem. 

\section{Polaron wave-function and spatial correlations}

\begin{figure*}[t]
    \centering
    \includegraphics[width=0.99\textwidth]{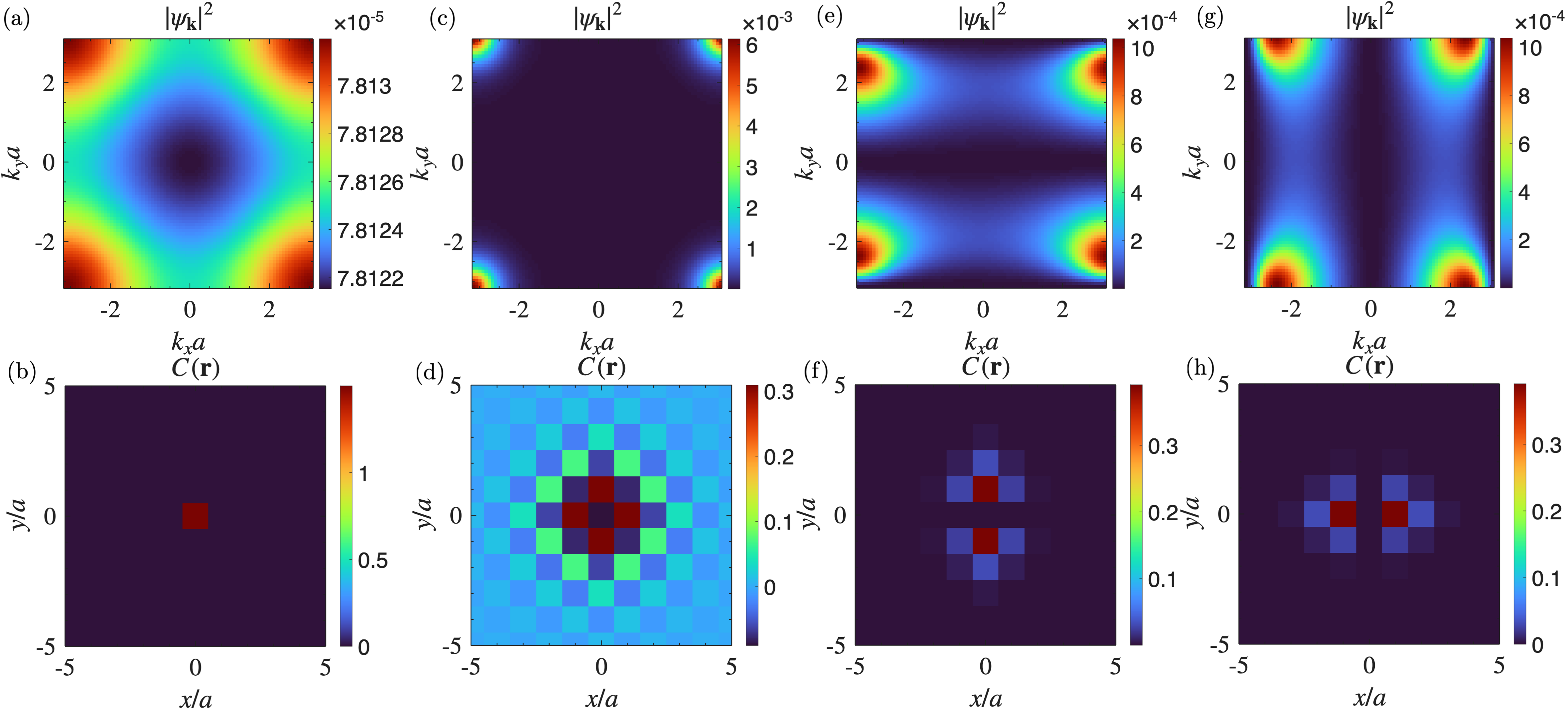}
    \caption{
    Momentum-space wave function (top row)  impurity-boson spatial correlation(bottom row)  for the four high-lying polaron states (a)-(b) Correspond to the highest polaron state for $U_I/t_b=2.0$. (c)-(d) show the wave function and impurity-boson spatial correlation for the second highest polaron branch for $U_I/t_b=2.0$. From (e)-(h) we show for two degenerate states featuring a dipole-like structure in real space for or $U_I/t_b=8.0$. These cases correspond to the eigenstates marked by the stars in Fig.~\ref{fig:spectral} and Fig.~\ref{fig:eigenstates}.
    }
    \label{fig:wf_high}
\end{figure*}

So far, we have understood the character of the polaron states from (a) the underlying two-body scattering problem, (b) the spectral function, and (c) the energy spectrum; however, to unveil their structure in both real and momentum space, we study the polaron wave function for the several polaron branches. This we show in Fig.~\ref{fig:wf_high}, where the top row illustrates the wave functions in momentum space and the bottom row gives the impurity-boson spatial correlation for several values of $U_I/t_b,$ marked by a star in Fig.~\ref{fig:spectral}.

Here, the spatial correlations between the impurity and the surrounding BEC are given by
\begin{equation}
C(\mathbf r_i-\mathbf r_j) = N \langle \psi| [\hat{n}_B(i) - n]\hat{n}_I(j) |\psi \rangle,
\end{equation}
where $\hat{n}_B(i) = \hat{b}_{\mathbf r_i}^\dagger \hat{b}_{\mathbf r_i}$ and $\hat{n}_I(i) = \hat{c}_{\mathbf r_i}^\dagger \hat{c}_{\mathbf r_i}$, 
$n = \langle \psi | \hat{n}_B(i) | \psi \rangle$ gives the total number of bosons at lattice site $i$. The correlation function $C(\mathbf r_i-\mathbf r_j)$ can be written in terms of the variational ansatz, for $\mathbf r_j=0$
\begin{equation}
C(\mathbf r_i) = 2\sqrt{\frac{n}{N}} \sum_{\mathbf k} \left[\mathcal R e \left(\varphi_0^* \psi_{\mathbf k} e^{i \mathbf k \cdot \mathbf r_i}\right)\right] 
+ \frac{1}{N}\left| \sum_{\mathbf k} \psi_{\mathbf k} e^{i \mathbf k \cdot \mathbf r_i}\right|^2,
\end{equation}
and characterizes the spatial distribution of bosons around the impurity, with the average density subtracted to isolate interaction-induced correlations. 

To understand the visible polaron branches we take $U_I/t_b=2.0.$ In Fig.~\ref{fig:wf_high}(a), we show the wave function  $\psi_{\mathbf k}$ of our variational approach in momentum space for the {\it 1st repulsive polaron state}, which has a minimum at $\mathbf{k} = 0$ and slowly increases at larger momentum. In real space, the impurity-boson correlation reveals that this state corresponds to the case where the impurity and a boson sit at the same site, shown in Fig.~\ref{fig:wf_high}(b). Here, the correlation function is peaked at $\mathbf r=0$ and is essentially structureless for $\mathbf r\neq 0.$
Thus, this state can be understood as reminiscent of the highest bound state.

For the {\it 2nd repulsive polaron branch}, we also take $U_I/t_b=2.0$ and we show that in Fig.~\ref{fig:wf_high}(c), 
the momentum-space wave function is concentrated near the corners of the Brillouin zone, 
$\mathbf{k} = (\pm\pi, \pm\pi)$, while it is strongly suppressed around $\mathbf{k} = 0$, 
in sharp contrast to the highest polaron state. This momentum-space structure translates  into a completely different correlation profile 
[Fig.~\ref{fig:wf_high}(d)]. In contrast to the {\it 1st repulsive polaron state}, the correlation function of the second polaron branch Fig.~\ref{fig:wf_high}(d) exhibits a qualitatively different behavior: the correlation is suppressed at $\mathbf r=0$ and instead develops a spatially extended pattern dominated by the first-neighbor correlations, and a correlation profile that extends over several lattice sites with a checkerboard pattern. Thus, we have that in this case, the polaron dresses by bosonic excitations with nontrivial spatial coherence.

To understand the {\it dark-impurity states}, we take $U_I/t_b=8.0$ and explore the corresponding states marked by the purple star in Fig.~\ref{fig:eigenstates}. We show in [Fig.~\ref{fig:wf_high}(e)], the momentum-space distribution remains concentrated near the Brillouin zone corners but exhibits a pronounced anisotropy, indicating a nontrivial superposition of large-momentum components. In real space [Fig.~\ref{fig:wf_high}(f)], the correlation function exhibits a nodal structure at $\mathbf r=0$, where it vanishes identically, and develops two off-center maxima along the $x$-direction. In contrast, the correlations remain strongly suppressed along the $y$-direction, resulting in a pronounced anisotropic profile. This pattern reflects a dipolar spatial structure aligned with the lattice axes, originating from destructive interference of momentum components near the edges of the Brillouin zone. 
This behavior sharply contrasts with the highest polaron states, which are dominated by on-site correlations and retain full lattice symmetry. The emergence of a node at $\mathbf r=0$ together with the directional dependence of the correlations indicates that these dark impurity states belong to a different symmetry sector, consistent with a $p$-wave–like character. 

A degenerate partner within the third-highest polaron branch exhibits the same nodal and anisotropic structure in both momentum and real space [Fig.~\ref{fig:wf_high}(g)–(h)]. In real space, the correlation function vanishes at $\mathbf r=0$ and develops two off-center maxima aligned along the $y$ direction, in contrast to the $x$-oriented state discussed above. This pair of states forms a symmetry-related doublet, consistent with a $p$-wave–like structure.

From the wave functions, we can then understand the dark impurity states. The noninteracting impurity state entering the spectral function is uniform in space and therefore transforms according to the fully symmetric $A_1$ irreducible representation of $C_{4v}$.

As a consequence, only eigenstates of the variational Hamiltonian that transform according to the same $A_1$ symmetry sector have a finite overlap with the noninteracting state and contribute to the spectral function. Eigenstates belonging to different irreducible representations are orthogonal to $|\psi_0\rangle$ by symmetry and therefore carry strictly zero spectral weight.

%The structure of the polaron wave functions reveals that the dark states exhibit a dipolar real-space profile with a node at $\mathbf{r} = 0$, together with an anisotropic distribution in momentum space concentrated near the edges of the Brillouin zone. These features are consistent with $p$-wave--like states transforming \luis{this is repeating the same idea as the paragraph before "A degenerate partner ... "}according to the two-dimensional $E$ representation of $C_{4v}$. In contrast, the bright polaron branches belong to the fully symmetric $A_1$ sector and therefore remain visible in the spectral response.

This symmetry classification provides a direct explanation for the vanishing spectral weight of the dark impurity states: their absence in the spectral function is not a consequence of the variational truncation, but rather follows from symmetry-imposed selection rules.

Therefore, the dark states identified in our calculations correspond to symmetry-forbidden quasiparticle excitations, which can only be accessed through probes sensitive to their internal structure rather than their overlap with the bare impurity. This symmetry argument states that the existence of dark impurity states persists beyond the Chevy ansatz.

Polaron states with vanishing spectral weight are not accessible in standard interferometric experiments. In contrast, quantum microscopy enables site-resolved probing of lattice systems~\cite{bakr2009quantum,gross2021quantum,gross2017quantum,mcdonald2019superresolution,schafer2020tools}. This has sparked renewed interest in polarons, including magnetic polarons~\cite{koepsell2019imaging,grusdt2018parton,nyhegn2023wave}, and makes our predictions directly testable in current ultracold-atom experiments.
%==========================================================
\section{Conclusions}

In summary, we have investigated the quasiparticle properties of a lattice polaron with strong on-site repulsion and tunable nearest-neighbor interactions using a Chevy-type variational framework. By combining the analysis of the spectral function with a direct inspection of the eigenvalue spectrum, we have shown that extended interactions qualitatively enrich the structure of polaron excitations. In addition to the conventional attractive and repulsive branches, we identified the emergence of additional repulsive polaron states whose visibility depends sensitively on the interaction strength.

A central result of our work is the identification of stable dark impurity states, which are orthogonal to the bare impurity and therefore carry vanishing spectral weight. Although spectroscopically dark, these states appear clearly in the eigenvalue spectrum and exhibit well-defined internal structure. In particular, we have shown that they display nontrivial symmetry properties, including dipolar real-space patterns, originating from the dominance of large-momentum components and their interference across the Brillouin zone. This highlights the importance of going beyond spectral observables to fully characterize quasiparticle excitations.

Our results show that nonlocal interactions give rise to hidden quasiparticle branches with distinct symmetry and spatial structure beyond conventional spectroscopic observables. It suggests anisotropic quasiparticle renormalization induced by extended interactions, where the resulting polaron dispersion is expected to become anisotropic, similar to dipolar polarons~\cite{Ardila2019Dipo,Kain2014,Volosniev2023,SanchezBaena2024}, leading to direction-dependent effective masses, anisotropic sound velocities, and enhanced finite-momentum dressing, and spatially anisotropic bound states (bipolarons).

These findings open new perspectives for probing polarons in engineered platforms with tunable interactions, such as dipolar gases and moiré materials, where momentum-selective or wave-function-resolved measurements could access otherwise hidden quasiparticle states. On the other hand, extended or nonlocal interactions can generate spatially structured many-body states beyond the simplest attractive- and repulsive-polaron picture. An interesting direction for future work is therefore to clarify whether anisotropic impurity bound states in lattice systems have a deeper connection to the many-body bound structures that appear in other long-range-interacting polaronic or dipolar-mixture systems~\cite{Bisset2021}.

The numerical data that supports the findings of this article are openly available~\cite{RamirezJuarez2026}. The codes are available upon request.

\section*{Funding information}
E. R-J, G. L-O and A. C-G acknowledge financial support from UNAM DGAPA PAPIIT Grant No. IA101325, Project CONAHCYT No. CBF2023-2024-1765 and PIIF25.

\begin{appendix}

\section{T-matrix approach}
\label{Tmatrix}
A useful benchmark of our results is the T-matrix approach, which resums repeated impurity-boson scattering processes in the ladder approximation~\cite{Rath2013}.  The starting point is the two-body scattering matrix
\begin{equation}
\mathcal T_\omega(\mathbf{p},\mathbf{p}') = V(\mathbf{p}-\mathbf{p}') + \frac{1}{N} \sum_{\mathbf{q}} 
V(\mathbf{p}-\mathbf{q}) \, \Pi_\omega(\mathbf{q}) \, \mathcal T_\omega(\mathbf{q},\mathbf{p}'),
\end{equation}
where $\Pi_\omega(\mathbf{q}) = [\omega - \epsilon^{(b)}_{\mathbf{q}} - \epsilon^{(c)}_{\mathbf{q}} + i\eta]^{-1}$ is the pair propagator. 

We then introduce the imaginary-time impurity Green's function 
\begin{gather}
 \mathcal G_c(\mathbf p,\tau)=-\langle T_\tau [\hat c_{\mathbf p}(\tau)\hat c_{\mathbf p}^\dagger(0)]\rangle,   
\end{gather}
which in momentum-energy space follows the Dyson's equation
\begin{gather}
  \mathcal G^{-1}_c(\mathbf p,\omega)=\omega-\epsilon_{\mathbf p}^{(0)}-\Sigma_c(\mathbf p,\omega). 
\end{gather}
Within the T-matrix formalism, the self-energy is given in terms of the impurity-boson scattering, thus, accounts for the exact two-body physics. For a zero-momentum impurity we take
\begin{gather}
 \Sigma_c(\mathbf p,\omega)=n\mathcal T_\omega(\mathbf 0,\mathbf 0),   
\end{gather}
and introduce the spectral function $$A(\omega) = -\,\mathrm{Im}\mathcal G_c(\omega).$$

Within the single-excitation (ladder) approximation, the $T$-matrix formalism captures the same impurity-boson scattering processes as the Chevy variational ansatz, and therefore yields consistent spectral properties for the polaron branches. In particular, the poles of the impurity Green's function correspond to the variational eigenstates that have finite overlap with the bare impurity. However, the variational approach additionally provides direct access to the full set of eigenstates and their wave functions, allowing us to identify states that are orthogonal to the bare impurity and therefore spectroscopically dark. In the present lattice system, these dark impurity states arise from the momentum structure of the interaction and the underlying lattice symmetries, which enforce selection rules on the overlap with the noninteracting impurity state. As a consequence, while the $T$-matrix correctly captures the visible quasiparticle branches, it does not reveal the full structure of the spectrum, highlighting the importance of combining both approaches to obtain a complete characterization of the polaron problem.

\subsection{Numerical details}
In contrast to the case of purely on-site interactions, where the $T$-matrix reduces to a scalar quantity due to the momentum independence of the interaction vertex, the presence of extended interactions in a lattice leads to a nontrivial momentum dependence $V(\mathbf{p}-\mathbf{p}')$. As a result, the $T$-matrix becomes a function of two independent momenta, $T(\omega;\mathbf{p},\mathbf{p}')$, and must be determined by solving a matrix equation. Upon discretizing the Brillouin zone using a grid of $N = N_x \times N_y$ momentum points, the integral equation for the $T$-matrix can be recast as
\begin{equation}
T_{ij}(\omega) = V_{ij} + \frac{1}{N} \sum_{k} V_{ik}\, \Pi_k(\omega)\, T_{kj}(\omega),
\end{equation}
where we have introduced the matrix elements $V_{ij} \equiv V(\mathbf{p}_i - \mathbf{p}_j)$ and the pair propagator $\Pi_k(\omega) = [\omega - \epsilon^{(b)}_{\mathbf{p}_k} - \epsilon^{(c)}_{\mathbf{p}_k} + i\eta]^{-1}$.

\begin{figure}[t]
\centering
\includegraphics[width=.45\columnwidth]{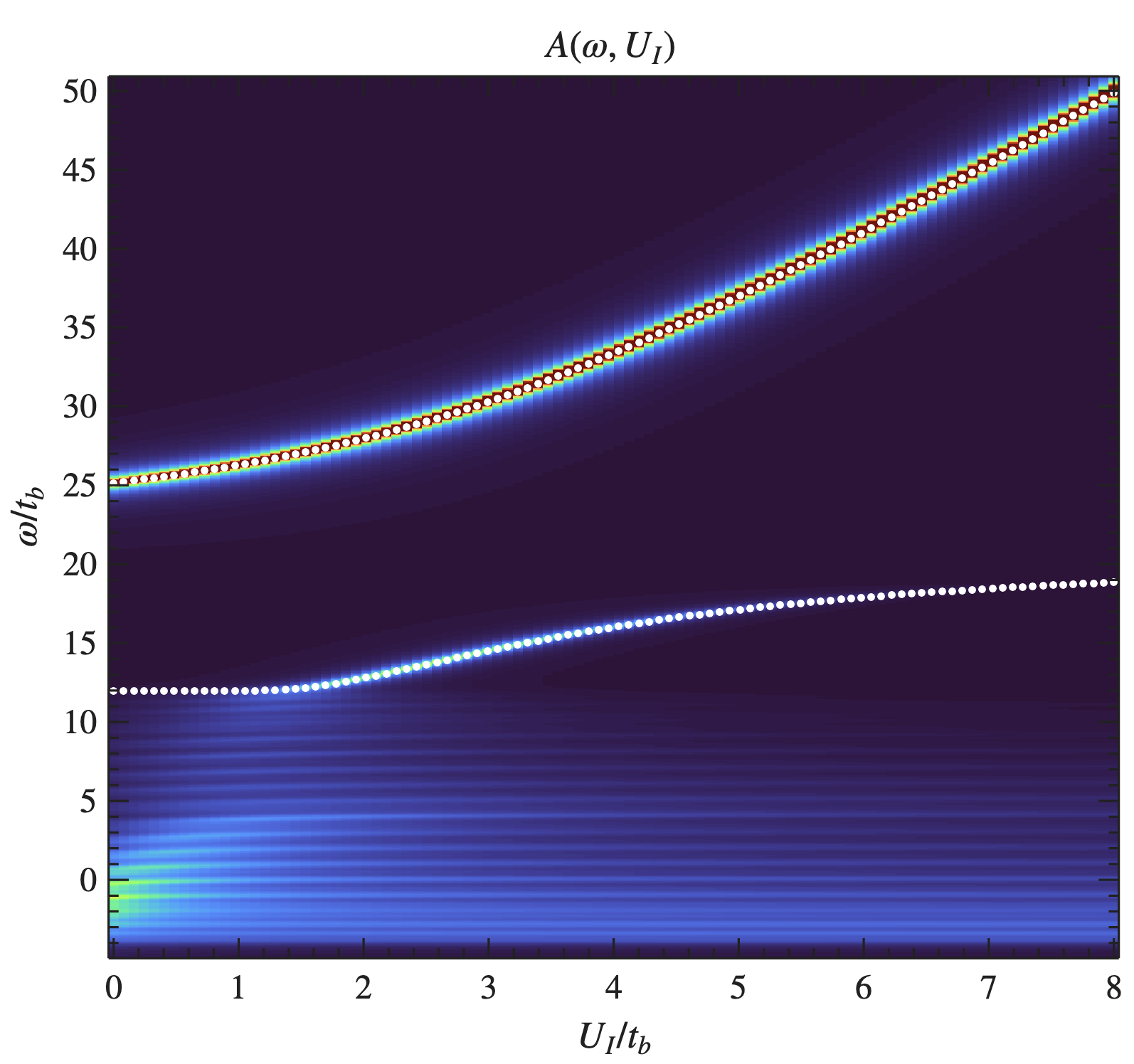}
\caption{
Spectral function $A(\omega,U_I)$ obtained from the T-matrix as a function of interaction strength $U_I/t_b$ and frequency $\omega/t_b$. The polaron branches extracted from the Chevy ansatz are shown by the white circles, including the 1st and 2nd repulsive polaron states. }
\label{fig:TM_vs_Chevy}
\end{figure}

This equation can be written compactly in matrix form as
\begin{equation}
\mathbf{T}(\omega) = \mathbf{V} + \mathbf{V}\,\mathbf{\Pi}(\omega)\,\mathbf{T}(\omega),
\end{equation}
where $\mathbf{V}$ is an $N \times N$ interaction matrix and $\mathbf{\Pi}(\omega)$ is a diagonal matrix with elements $\Pi_k(\omega)$. Rearranging, we obtain a linear system
\begin{equation}
\mathbf{T}(\omega) =\left[\mathbf{I} - \mathbf{V}\,\mathbf{\Pi}(\omega)\right]^{-1} \mathbf{V},
\end{equation}
 The numerical solution thus involves inverting an $N \times N$ matrix for each $\omega$, making the computation significantly more demanding than in the local-interaction case, where the $T$-matrix reduces to a scalar equation. This increased complexity reflects the fact that extended interactions couple all momentum states, leading to a fully nonlocal scattering problem in momentum space.

 In Fig.~\ref{fig:TM_vs_Chevy} we show the spectral function obtained from the T-matrix approach which agrees with the spectral function in Fig.\ref{fig:spectral} from the Chevy's ansatz. The white dots correspond to the two highest polaron states obtained from the Chevy's ansatz. In this case, we employ a $N_x\times N_y=576$ sites and an imaginary width of $\eta/t_b=0.2$

\end{appendix}
\bibliography{references_Lattice}

%%%%%%%%%% END TODO: BIBLIOGRAPHY

\end{document}